\documentclass[pra,aps,superscriptaddress,floatfix,tightenlines,showpacs,twocolumn]{revtex4}

\usepackage{graphicx}
\usepackage{dcolumn}   
\usepackage{bm}        
\usepackage{amssymb}   
\usepackage{amsmath}
\usepackage{url,color}
\usepackage{layout}
\usepackage{epsfig}
\usepackage{graphicx}
\usepackage{epstopdf}
\usepackage{booktabs}
\usepackage{float,CJK}
\usepackage[hyperindex,breaklinks]{hyperref}
\newcommand{\ket}[1]{|#1\rangle}                  
\newcommand{\bra}[1]{\langle #1 |}     

\begin{document}
\title{Optimizing Quantum Adiabatic Algorithm}
\begin{CJK*}{UTF8}{gbsn}

\author{Hongye Hu(扈鸿业)}
\affiliation{International Center for Quantum Materials, School of Physics, Peking University, Beijing 100871, China}
\author{Biao Wu(吴飙)}
\email{wubiao@pku.edu.cn}
\affiliation{International Center for Quantum Materials, School of Physics, Peking University, Beijing 100871, China}
\affiliation{Collaborative Innovation Center of Quantum Matter, Beijing, China}
\affiliation{Wilczek Quantum Center, College of Science, Zhejiang University of Technology,  Hangzhou 310014, China}

\begin{abstract}
In  quantum adiabatic algorithm, as the adiabatic parameter $s(t)$ changes
slowly from zero to one with finite rate, a transition to excited states inevitably occurs
and this induces an intrinsic computational error. We show that this computational error
 depends not only on the total computation time $T$ but also on
the time derivatives of the adiabatic parameter $s(t)$ at the beginning and the end of evolution. Previous work (Phys. Rev. A \textbf{82}, 052305) also suggested this result. 
With six typical paths, we systematically demonstrate how to optimally design an adiabatic path to
reduce the computational errors. Our method has a clear physical picture and also explains the pattern of computational error. In this paper we focus on quantum adiabatic search algorithm
although our results are general. 
\end{abstract}

\maketitle
\end{CJK*}

\section{Introduction}

Quantum adiabatic algorithm was proposed in 2000 by Farhi {\it et al.}\cite{Farhi0} as an alternative
paradigm of quantum computing to quantum circuit algorithm~\cite{QuantumComputationBook}.  
It works by constructing a time-dependent Hamiltonian that evolves slowly from the initial Hamiltonian to the problem Hamiltonian. The ground state of the initial Hamiltonian is easy to find and the answer of the intent problem is encoded in the problem Hamiltonian. It is ensured by the quantum adiabatic theorem that if the Hamiltonian evolves slowly enough, the system will stay in the ground state and evolve into  the ground state of the problem Hamiltonian. 
When it is applied to the search problem, the quantum adiabatic algorithm has been shown to be $O(\sqrt{N})$~\cite{Roland, RezakhaniPRL}, which is as powerful as Grover's 
algorithm~\cite{Grover} and is quadratic speedup over the classical search algorithm. 
In general, the adiabatic algorithm has been shown to have the potential to solve NP-hard 
problem~\cite{Farhi2, Zhuang}. 

In addition to its speed up against classical computing, the quantum adiabatic algorithm also has capacity to remain robust against environment noise~\cite{Preskill}. Some practical architectures for quantum adiabatic algorithms were proposed~\cite{Seth}; in 2013 D-Wave company claimed that they built a quantum computer based on quantum adiabatic algorithm~\cite{DWAVE}.

 
In quantum adiabatic algorithm, there are two types of computational errors. One is the extrinsic
error, which is caused by the environment. The other is the intrinsic error:
as the algorithm has to be run in a finite computation time $T$, the adiabatic Hamiltonian
must change in a finite rate and this inevitably will induce transition to excited states and cause computational error. The intrinsic computational error depends entirely on how the adiabatic path $s(t)$ is chosen. The most popular choice so far is the linear path, $s(t)=t/T$.  Other choices were proposed in literature~\cite{Rezakhani}. People has also tried to optimize the adiabatic path $s(t)$
using geometrization~\cite{RezakhaniPRL}.

{ According to the  hierarchical theory of quantum adiabatic evolution~\cite{Qi},   
the intrinsic error depends crucially on  the time derivatives of $s(t)$ at the
beginning and the end of the evolution.  This fact was also pointed out in 
Ref.  \cite{Rezakhani,Lidar}.  }
In this work we choose six typical adiabatic paths $s(t)$ 
to systematically demonstrate how to optimally control these
time derivatives to reduce the computational error. Also we give a physical picture about the origin of computational error's oscillation. We focus on the quantum adiabatic
search algorithm and numerically compare the computational errors for different adiabatic paths.
We find that the cubic path is the best among the six chosen path and 
it can reduce the error by orders of magnitude compared to the popular linear path and
sinusoidal path. 

Our paper is organized as follows: for the sake of self-containment, we first briefly introduce 
the quantum adiabatic algorithm, in particular, the quantum adiabatic search algorithm, and 
the hierarchical theory of quantum adiabatic evolution, respectively, in Sections II and III. 
In Section IV, we show that the Hamiltonian for the quantum adiabatic search algorithm 
can be reduced to a spin-1/2 Hamiltonian and apply the hierarchical theory. In Section V, 
six typical adiabatic paths are chosen and they are categorized into three groups. Our main
numerical results are shown in Section VI. We finally conclude in Section VII.

\section{Quantum Adiabatic Search Algorithm}
The adiabatic quantum computation was first introduced  in 2000~\cite{Farhi0} based
on the quantum adiabatic theorem~\cite{BornFock}.  In a quantum adiabatic algorithm,
the solution of a problem is encoded into the ground state of the problem Hamiltonian $H_{p}$.
An initial Hamiltonian $H_{b}$ is chosen so that its ground state can be
easily found and set up. The total adiabatic Hamiltonian is constructed by
 linking the initial Hamiltonian with the problem Hamiltonian with a path as follows
 \begin{equation}
H_s(t)=(1-s(t))H_{b}+s(t)H_{p}\,,
\label{H}
\end{equation}
where  $s(t)\in [0,1]$ is the adiabatic parameter.  
The system is prepared in the ground state of $H_{b}$.
As $s$ changes slowly from zero to one, the quantum adiabatic theorem ensures
that the system stays in the ground state of $H_s$ and eventually arrives at
the ground state of $H_{p}$, the solution.

In the  search problem, the task is to locate $M$ marked items out of $N$ randomly arranged items.
On a quantum computer,  we use a set of orthonormal basis $|1\rangle, |2\rangle, ...,|N\rangle$ to denote the $N$ unsorted items. The problem Hamiltonian can be constructed as~\cite{Roland}
\begin{equation}
H_{p}=1-\sum_{m\in \mathcal{M}}|m\rangle \langle m|,
\end{equation}
where $\mathcal{M}$ is the set of marked items. This Hamiltonian is the projection operator to a subspace orthogonal to the subspace spanned by $\{|m\rangle\}_{m\in\mathcal{M}}$; its ground state can be any state from subspace $\{|m\rangle\}_{m\in\mathcal{M}}$. If we choose the initial state to be equally contributed from the orthonormal basis as $|\psi_0\rangle=\dfrac{1}{\sqrt{N}}\sum_{i=1}^{N}|i\rangle$, then the initial Hamiltonian is
\begin{equation}
H_{b}=1-\ket{\psi_0}\bra{\psi_0}=1-\dfrac{1}{N}\sum_{i,j} |i\rangle \langle j|.
\end{equation}
Thus the quantum adiabatic Hamiltonian for search is
\begin{equation}
H_{s}=1-\dfrac{1-s(t)}{N}\sum_{i,j}|i\rangle \langle j|-s(t)\sum_{m\in \mathcal{M}}|m\rangle \langle m|\,.
\end{equation}

Note that the unit of Hamiltonian which rules the quantum computation depends on what kind of system is used for realization. Suppose a system has a characteristic time $\tau$; the unit of Hamiltonian is $\hbar/\tau$. For convenience, in the following derivation, we set $\tau=\hbar=1$.

In the above Hamiltonian, if $s(t)$ changes from zero to one infinitely slowly, the system will stay
strictly in the ground state and the solution can be found without any error if there
is no environment noise. This is dictated by the quantum adiabatic theorem.
However, we want to know the solution as fast as possible. This means that $s$ has
to change from zero to one in a finite time $T$, causing a small transition to excited states. 
At the end, there is an inevitable error in the solution. For the search problem,
we define the intrinsic computational  error as~\cite{Rezakhani}
\begin{equation}
\delta =1-\langle\psi(T)|\hat{P}|\psi(T)\rangle\,,
\end{equation}
where the projection operator $\hat{P}$  is defined as
\begin{equation}
\hat{P}=\sum_{m\in \mathcal{M}}|m\rangle \langle m|\,.
\end{equation}

The main purpose of this work is to reduce the computational error $\delta$ by optimizing
the adiabatic path $s(t)$. The simplest and also the most popular choice is the linear
path $s=t/T$. This is almost the worst among the easy choices, which includes sinusoidal
path. According to the newly developed hierarchical theory of quantum adiabatic
evolution~\cite{Qi}, the error $\delta$ depends crucially how the time derivatives of $s(t)$
at the beginning and the end of the adiabatic evolution. In this work, several adiabatic paths $s(t)$ are
designed and the errors caused by these paths are computed and compared to the error
by the linear path. The error can be reduced by orders of magnitude.

\section{Hierarchical Theory of Quantum Adiabatic Evolution}

\begin{figure}[htbp]
\centering
\includegraphics[width=0.99\linewidth]{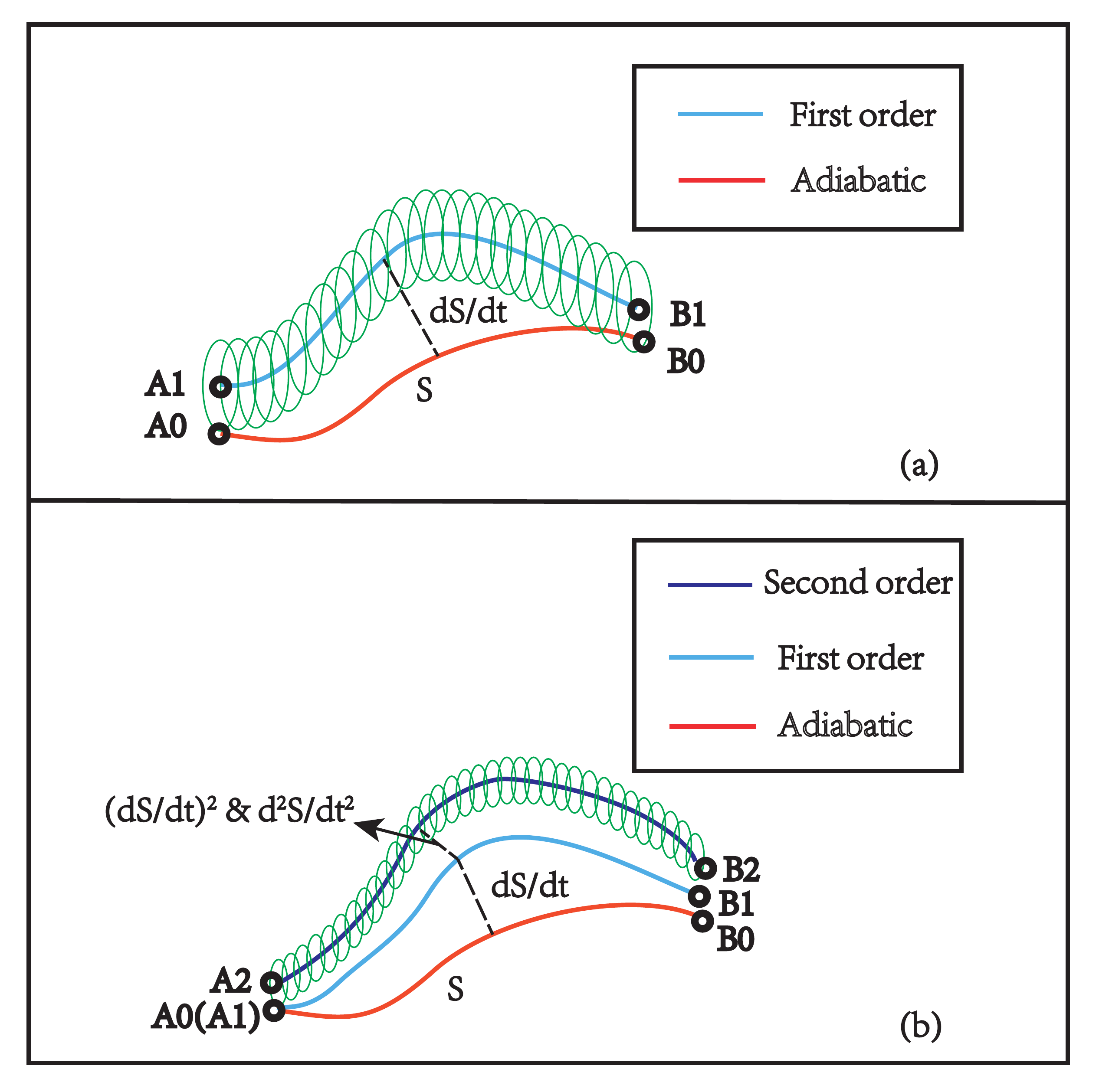}
\caption{\label{proj}(color online)Adiabatic evolution trajectories in the projective Hilbert space.
The red lines are the zeroth-order trajectories, that is, the trajectories follow strictly 
the quantum adiabatic theorem.  The blue lines are the first-order trajectories, which
are shifted from the zeroth-order  ones by a small amount proportional to $\dot{s}$.
The black lines are the second-order trajectories. When $\dot{s}$ is not zero at $t=0$,
the system will oscillate around the first-order trajectory (the green line in (a)).
Similarly, when $\ddot{s}$ is not zero at $t=0$,  the system will oscillate around
the second-order trajectory (the green line in (b)).}
\end{figure}

The quantum adiabatic theorem was proved in 1928 by Born and Fock~\cite{BornFock}.
This theorem ensures that a system starting in the ground state will stay in the ground state
when the adiabatic parameter $s$ changes slowly. However, this is mathematically true
only when the changing rate of $s$ is infinitesimally small. In any practical situation, for example,
quantum adiabatic computing,  the adiabatic parameter $s$ has to change with a small but
finite rate, this will cause a small transition to excited states, resulting a deviation from the
quantum adiabatic theorem.  In Ref. \cite{Qi},  a hierarchical theory was developed to compute
the deviation order by order.

The results are schematically illustrated in Fig. \ref{proj} with trajectories in the projective 
Hilbert space (the overall phase is not important)~\cite{Fu1,Fu2,Niu}.  
At the zeroth order, the system follows the
trajectory dictated by the quantum adiabatic theorem. At the first order, the system oscillates
with a small amplitude around a trajectory slightly shifted from the zeroth-order trajectory.
The small shift is proportional to $\dot{s}$, the first-order time derivative of $s(t)$; the small oscillating
amplitude is determined by $\dot{s}$ at $t=0$.  This is illustrated in Fig. \ref{proj}(a).

To reduce and also to better control the computational error $\delta$,
we can set $\dot{s}=0$ at the beginning. In this case, there are no oscillations and
the system will follow a smooth first-order trajectory. As the shift between the first-order
trajectory and the zeroth-order trajectory is proportional to $\dot{s}$,  we can reduce
the error $\delta$ in the first order to zero by choosing a $s(t)$ such that $\dot{s}=0$ at $t=T$.
For such a path $s(t)$, the error $\delta$ is of the second-order, determined by $\ddot{s}$, the
second-order time derivative of $s(t)$.  To further reduce the error, we can repeat
the above procedure by choosing a path $s(t)$ such that $\ddot{s}=0$ at $t=0,T$.
This strategy is very successful as we shall see in the following sections. 

 We note here that the crucial role played by the time derivatives was also 
pointed out in Ref. \cite{Rezakhani,Lidar} with a different approach. 
\section{Reduced Search Hamiltonian and Analytical Results}
\begin{figure}[H]
\centering
\includegraphics[width=1\linewidth]{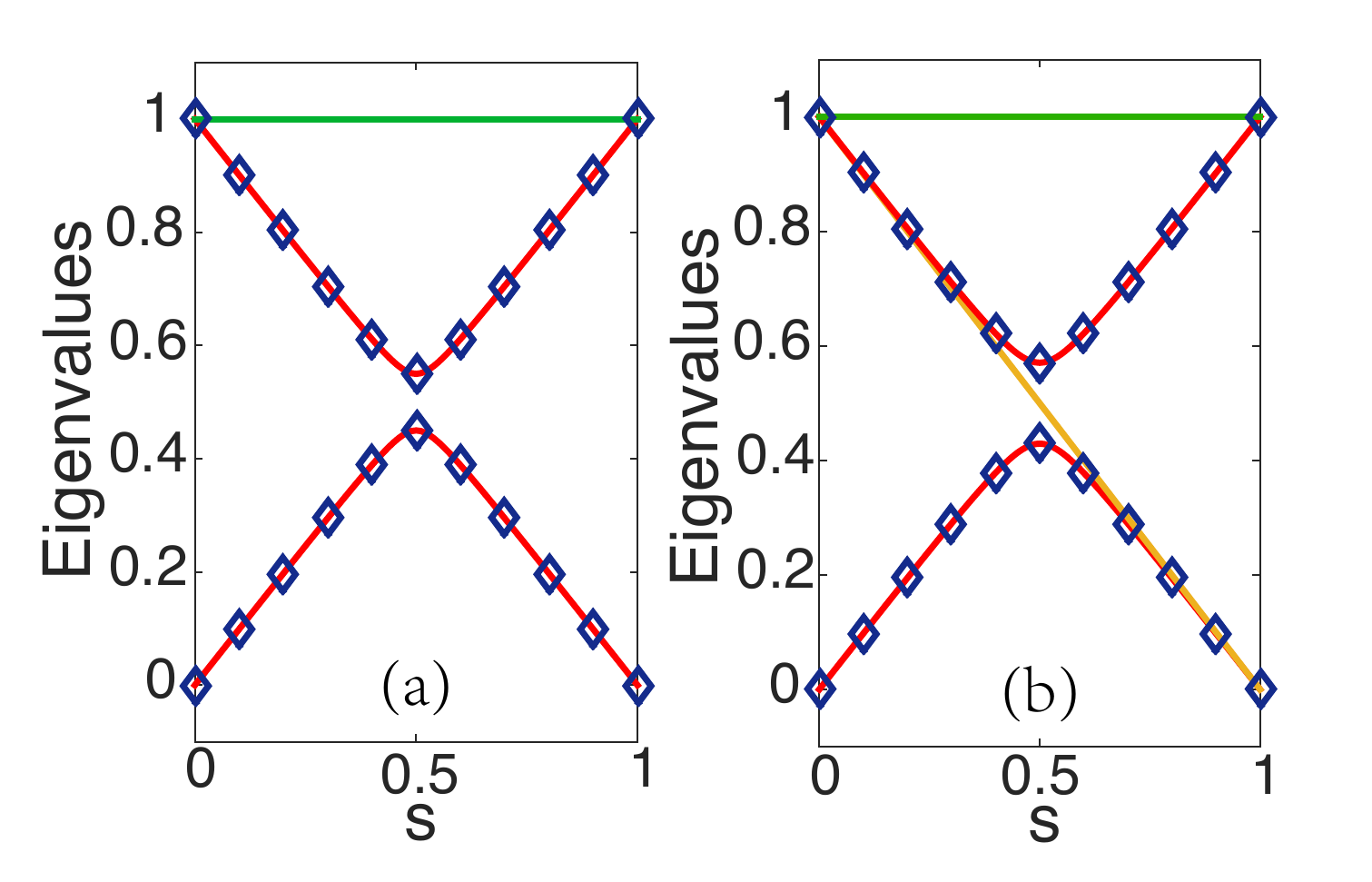}
\caption{(color online)Eigenvalues of the search Hamiltonian.  (a) $N=100$, $M=1$; (b) $N=100$, $M=2$.
The solid lines are for $H_s$ and the diamonds for $\widetilde{H}_s$. The time unit 
depends on the system used to realize our algorithm. Suppose the system has a characteristic time $\tau$, then the unit of energy is $\hbar/\tau$. In this article, we have set $\hbar = \tau = 1$.}
\label{eigenhigh}
\end{figure}

Let us come back to the search Hamiltonian and see how the hierarchical theory
outlined in the last section can be applied successfully to this problem.  
Because of the permutation symmetry of the search Hamiltonian $H_s$, 
for the given initial state, the quantum state at any given time has the form~\cite{VDam2}
\begin{equation}
|\psi\rangle=\frac{\psi_{u}}{\sqrt{N-M}}\sum_{u\notin\mathcal{M}}|u\rangle +
\frac{\psi_{m}}{\sqrt{M}}\sum_{m\in \mathcal{M}}|m\rangle\,.
\end{equation}
The Schr\"odinger equation governing the search algorithm becomes 
\begin{equation}
i\dfrac{\partial}{\partial t}\begin{pmatrix}\psi_{u}\\ \psi_{m}\end{pmatrix}=\widetilde{H}_s\begin{pmatrix}\psi_{u}\\ \psi_{m}\end{pmatrix},
\label{schr}
\end{equation}
with 
\begin{equation}
\widetilde{H}_s=\begin{pmatrix}r(1-s(t))+s&-\sqrt{r(1-r)}(1-s(t))\\-\sqrt{r(1-r)}(1-s(t)) &(1-r)(1-s(t))\end{pmatrix}
\end{equation}
where $r=M/N$. In other words, the search Hamiltonian $H_s$ is reduced to a spin-1/2
Hamiltonian $\widetilde{H}_s$. Fig. \ref{eigenhigh} shows the eigenvalues of both 
Hamiltonians, $H_s$ and $\widetilde{H}_s$. It is clear that two eigenvalues 
of $\widetilde{H}_s$ (diamonds) are identical to two of the eigenvalues of $H_s$ (red solid lines). 
Due to the permutation symmetry,  all other eigenstates of $H_s$ (non-red solid lines) 
do not participate in the dynamical evolution when $s$ changes from zero to one. \\


To apply the hierarchical theory in  Ref.\cite{Qi}, 
we reformulate our problem in the projective Hilbert space~\cite{Fu1,Fu2,Niu}. 
We rewrite the state as
\begin{equation}
|\psi_{2}\rangle=\begin{pmatrix}\psi_{u}\\ \psi_{m}\end{pmatrix}=\dfrac{\lambda_1}{\sqrt{2}}\begin{pmatrix}1\\ i\end{pmatrix}+\dfrac{\lambda_2}{\sqrt{2}}\begin{pmatrix}1\\-i\end{pmatrix}\,,
\end{equation}
and define $p=\arg(\lambda_2)-\arg(\lambda_1)\in [0,2\pi]$,  $q=|\lambda_2|^2\in[0,1]$. 
The dynamics of our system in the projective Hilbert space can be completely determined by
$p$ and $q$.  In terms of $p$ and $q$, the classical Hamiltonian becomes
\begin{eqnarray}
\widetilde{H}_c&=&\dfrac{1}{2}+\sqrt{q(1-q)}\cos(p)(2r+2s-2rs-1)\nonumber\\
&&-2\sqrt{q(1-q)}\sin(p)\sqrt{r(1-r)}(1-s)\,,
\end{eqnarray}
where $\widetilde{H}_c = \langle \psi_{2}|\widetilde{H}_s|\psi_{2}\rangle$.\newline

The ground state of the reduced search Hamiltonian is a fixed point in the projective Hilbert space
where the overall phase is removed. The fixed point is given by
\begin{equation}
\left\{
\begin{array}{ll}
\bar{q}=\dfrac{1}{2}\\
\bar{p}=
\left\{
\begin{array}{ll}\pi -\arctan(\dfrac{2\sqrt{(1-r)r}(1-s)}{2r+2s-2rs-1}).~~(s\geq \dfrac{1-2r}{2-2r})\\
\\
\arctan(\dfrac{2\sqrt{(1-r)r}(1-s)}{2rs-2r-2s+1}).~~(s< \dfrac{1-2r}{2-2r})
\end{array}
\right.
\end{array}
\right.
\end{equation}
where we have assumed that $r\ll 1$.
When $s$ changes infinitesimally slowly, this fixed point traverses the adiabatic trajectory 
(the red lines in Fig. \ref{proj}).  When $s$ changes with a finite but small rate, 
the actual dynamics will deviate from the adiabatic trajectory of the fixed point. 
The averaged first-order and second-order deviations are\cite{Qi}
\begin{enumerate}
\item \textit{First Order Deviation}\\
\begin{equation}
\begin{pmatrix}A_1\\B_1\end{pmatrix}=\begin{pmatrix}0\\ \dfrac{\dot{s}\sqrt{r(1-r)}}{\lambda^{3/2}}\end{pmatrix}
\end{equation}
\item \textit{Second Order Deviation}\\
\begin{equation}
\begin{pmatrix}A_2\\B_2\end{pmatrix}=\begin{pmatrix}2\sqrt{r(1-r)}\dfrac{3\dot{s}(r-1)(2-4s)-\lambda \ddot{s}}{\lambda^3}\\ 0 \end{pmatrix}
\end{equation}
\end{enumerate}
where $\lambda=1+4(r-1)s-4(r-1)s^2$.  $A_i$ is the ith order deviation of p, and $B_i$ is the ith order deviation of q. Detailed derivation of ($A_{i}, B_{i}$) can be found in the appendix. The first-order deviation 
as depicted by the blue lines in  Fig. \ref{proj} is proportional to $\dot{s}$. 
The second-order deviation is
 depicted by the black line in  Fig. \ref{proj}(b) and it is proportional to $\ddot{s}$ if
 $\dot{s}$ is zero.

\section{Adiabatic Paths}
We will show in the last two sections that 
the time derivatives of $s$ at the beginning and end 
play a crucial role in determining
the computational errors.  We define these time derivatives as
\begin{equation}
c_n\equiv\dfrac{d^{n}s}{dt^n}(0)~~\,,~~~~~~d_n\equiv\dfrac{d^{n}s}{dt^n}(T)\,.
\end{equation}
With $c_n$ and $d_n$, we categorize the evolution paths
in the following ways: the path is called $n$th-order if all of its $c_m$ and $d_m$ for $m\le n$ 
are zero while either $c_n$ or $d_n$ are not zero. The zeroth-order path has 
either $c_1\neq 0$ or $d_1\neq 0$. We propose the following six adiabatic paths to illustrate  
our ideas:
\begin{enumerate}
\item Linear path
\begin{equation}
s_1(t)=
\left\{
\begin{array}{lll}
0~~~t<0\\
\dfrac{t}{T}~~~0\leq t<T\\
1~~~t\geq T
\end{array}
\right.
\end{equation}
\item Sinusoidal path
\begin{equation}
s_2(t)=
\left\{
\begin{array}{lll}
0~~~t<0\\
\sin(\frac{\pi t}{2T})~~~0\leq t<T\\
1~~~t\geq T
\end{array}
\right.
\end{equation}
\item Square path
\begin{equation}
s_3(t)=
\left\{
\begin{array}{lll}
0~~~t<0\\
3(\frac{t}{T})^2-2(\frac{t}{T})^3~~~0\leq t<T\\
1~~~t\geq T
\end{array}
\right.
\end{equation}
\item Sinusoidal square path
\begin{equation}
s_4(t)=
\left\{
\begin{array}{lll}
0~~~t<0\\
\sin^2(\frac{\pi t}{2T})~~~0\leq t<T\\
1~~~t\geq T
\end{array}
\right.
\end{equation}
\item Sinusoidal cubic path
\begin{equation}
s_5(t)=
\left\{
\begin{array}{lll}
0~~~t<0\\
\sin^3(\frac{\pi t}{2T})~~~0\leq t<T\\
1~~~t\geq T
\end{array}
\right.
\end{equation}
\item Cubic path
\begin{equation}
s_6(t)=
\left\{
\begin{array}{lll}
0~~~t<0\\
6(\frac{t}{T})^5-15(\frac{t}{T})^4+10(\frac{t}{T})^3~~~0\leq t<T\\
1~~~t\geq T
\end{array}
\right.
\end{equation}
\end{enumerate}

Among these six paths, linear path and sinusoidal path are of zeroth order;  
the first-order paths are square path, sinusoidal square path, and sinusoidal cubic path; 
cubic path is second order.  As we shall see, the higher-order path 
can lead to smaller computational error by orders of magnitude. 
Note that only linear path and sinusoidal path of those six paths were studied before.~\cite{Rezakhani}

\begin{figure}[htbp]
\centering
\includegraphics[width=0.95\linewidth]{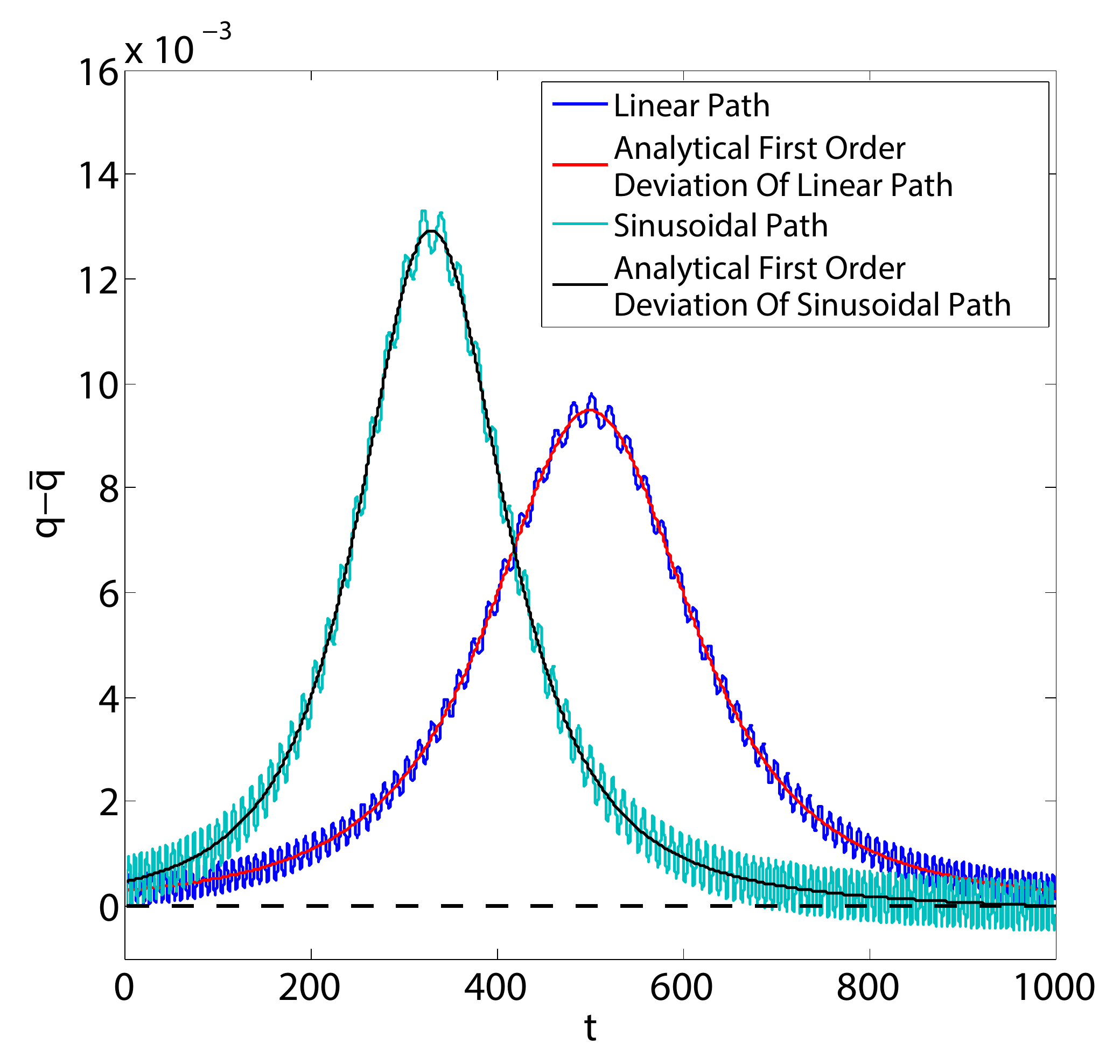}
\caption{(color online) Deviations from the ground state as a function of time for the linear path 
and the sinusoidal path. The analytical results for the first-order deviation are 
plotted for comparison.   ($N=10$, $M=1$, $T=1000$) \label{first}}
\end{figure}
\section{Numerical Results}
The Schr\"odinger equation (\ref{schr}) is solved numerically for the six adiabatic paths
for various values of $T$.  The results are analyzed, explained, and compared in this section. 
They convincingly show that higher-order paths can reduce computational errors 
by orders of magnitude.

Fig. \ref{first} shows how the deviation from the 
ground state (fixed point) changes with time for linear path and sinusoidal path. 
It is clear from this figure that the deviations oscillate around the first-order analytical 
results. And the amplitude of this oscillation is conserved, which is an result of conservation of action\cite{Dirac}. This is expected from the hierarchical theory (see Fig. \ref{proj})\cite{Qi} as
both linear path and  sinusoidal path are zeroth-order path with nonzero $c_1$.  
It is interesting to compare the results for linear path and sinusoidal 
path. The deviation for the sinusoidal path is larger than
that for the linear path in the middle of the evolution; however, 
the computational error (i.e., the deviation at $t=T$) 
appears slightly smaller for  sinusoidal path.  The reason is that $\dot{s}$ decreases 
smoothly toward zero for the sinusoidal path. Although the reduction of the computational
error is not much, it already shows the possibility to reduce the computational error by optimally
designing $s(t)$ for a given $T$.  This kind of reduction can be of order of magnitude
when we choose a path of higher-order as we shall see next.

\begin{figure}[htbp]
\centering
\includegraphics[width=0.95\linewidth]{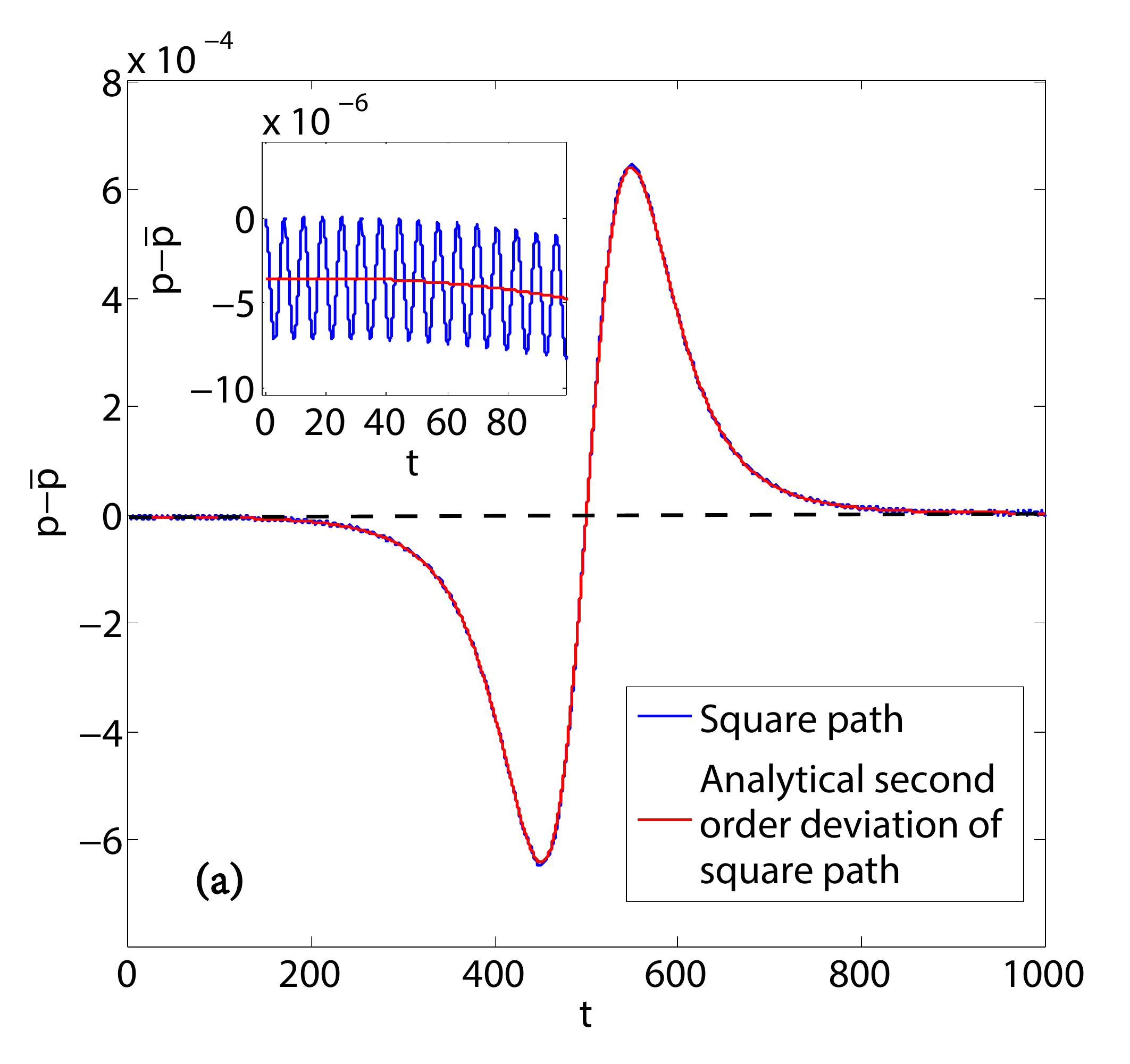}\\
\includegraphics[width=0.97\linewidth]{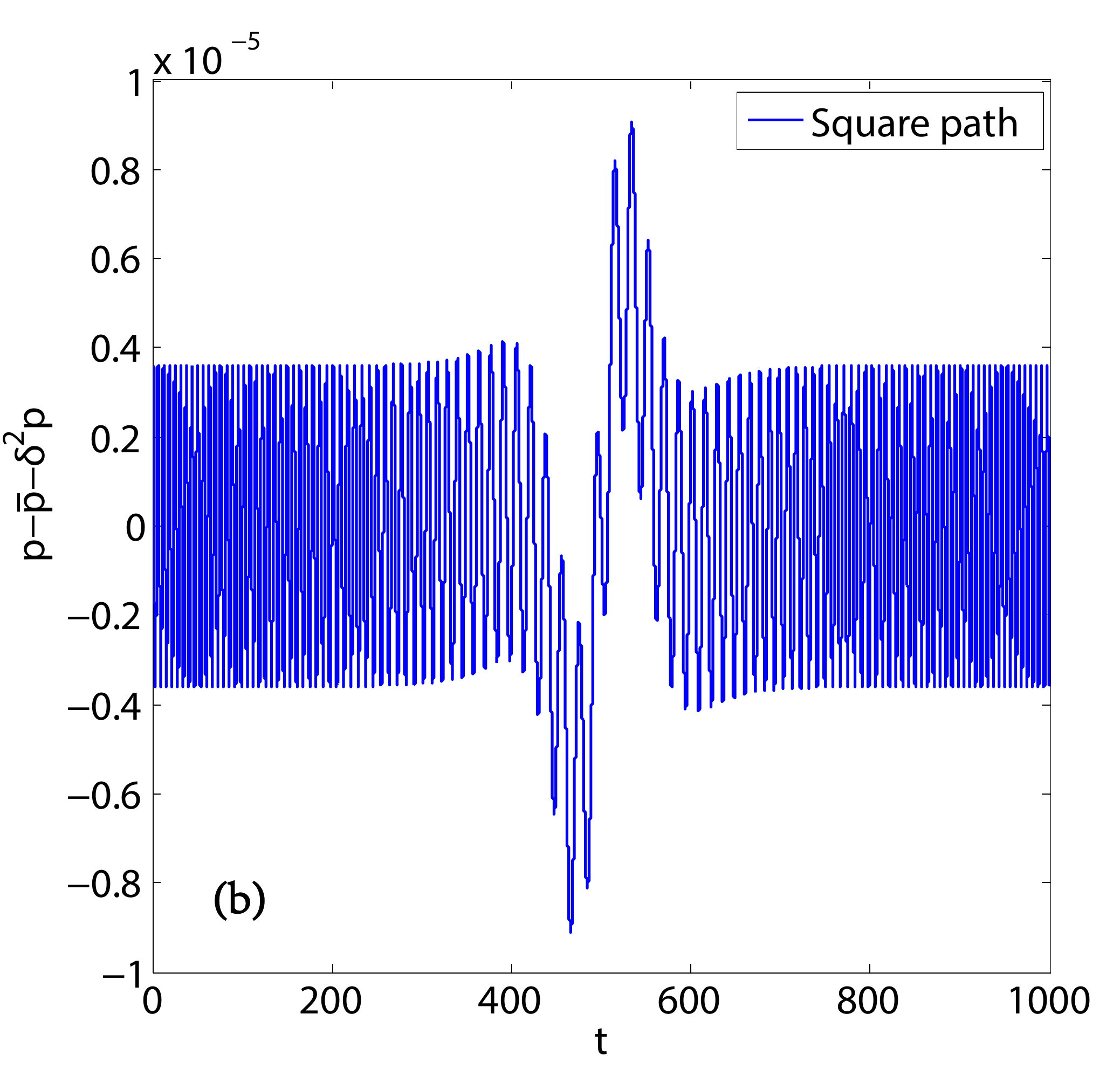}
\caption{(color online) (a) Deviation from the ground state as a function of time for the square path. 
The analytical results for the second-order deviation are 
plotted for comparison.  (b) Difference between numerical result of p 
and its second order analytical result as a function of time. ($N=10$, $M=1$, $T=1000$)}
\label{second}
\end{figure}


\begin{figure}[htbp]
\centering
\includegraphics[width=0.95\linewidth]{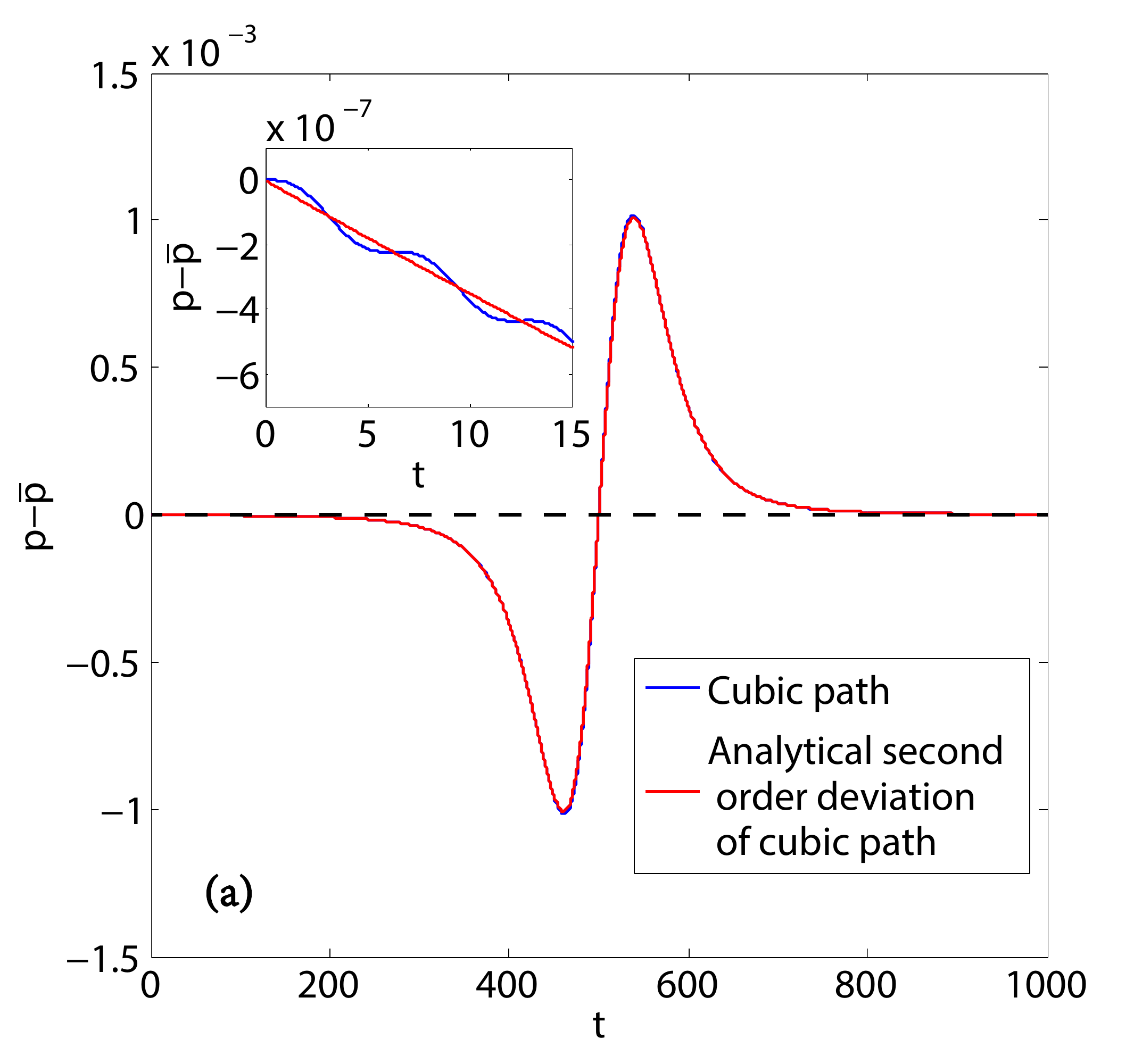}\\
\includegraphics[width=0.95\linewidth]{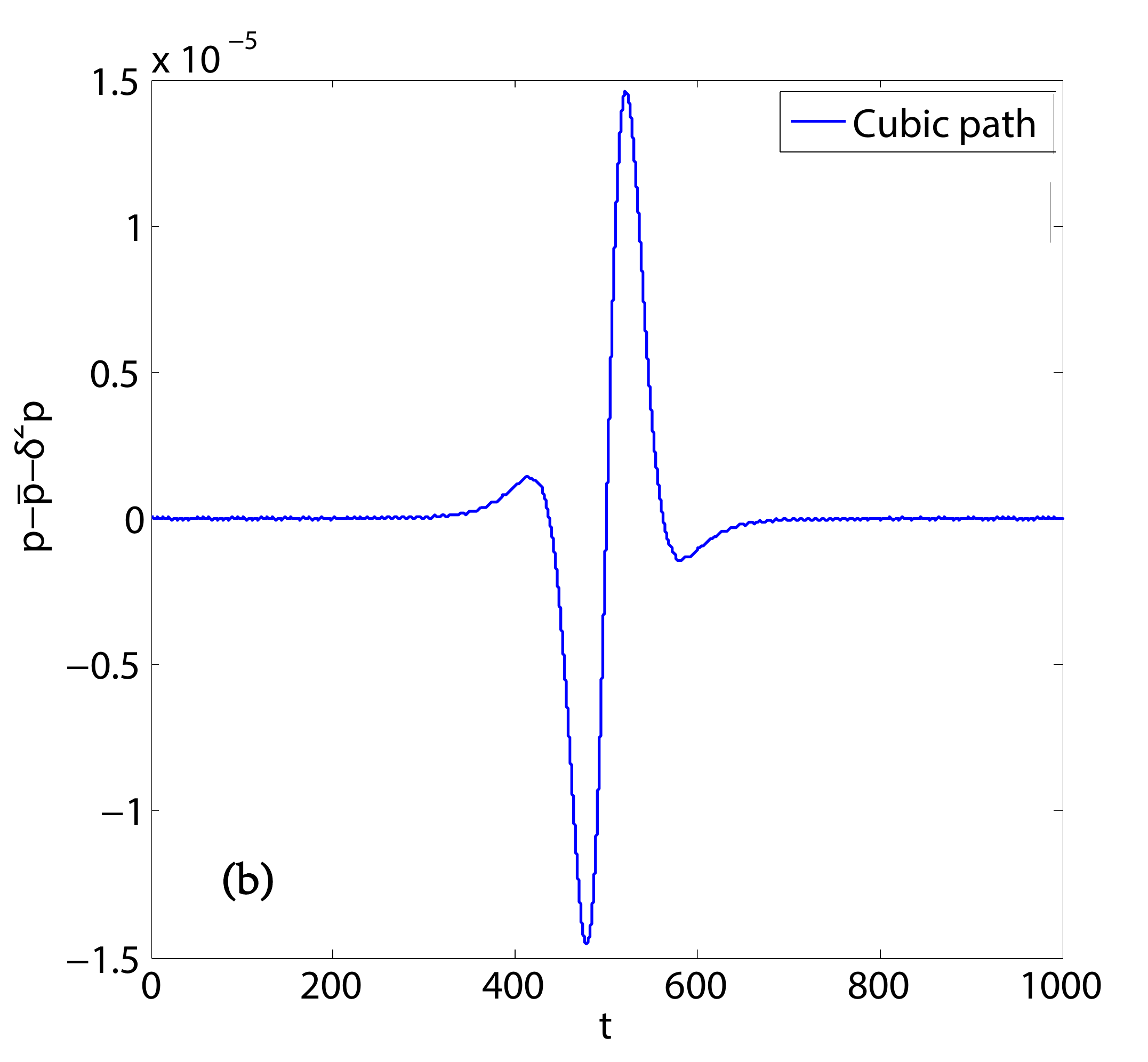}
\caption{(color online) (a) Deviation from the ground state as a function of time for the square path. 
The analytical results for the second-order deviation are 
plotted for comparison.  (b) Difference between numerical result of p and 
its second order analytical result as a function of time. 
($N=10$, $M=1$, $T=1000$)\label{cubicp}}
\end{figure}


We have designed three first-order paths, square path, sinusoidal square path,
and sinusoidal cubic path whose $c_1$ and $d_1$ are zero. As an example, 
our numerical results for square path is plotted in Fig.\ref{second}. 
The results in  Fig.\ref{second}(a) are  deviation from the ground state. 
As the first-order derivative $c_1$ and $d_1$ are zero, the deviation
is of second order and it oscillates around the analytical second-order
deviation (see the inset of Fig.\ref{second}(a)). To see these oscillations
more clearly, we have plotted the difference between our numerical results
and the analytical second-order deviation in Fig.\ref{second}(b)), 
where the oscillation pattern is seen to have a kink in the middle of the evolution.  
Since the second order derivatives $c_2$ and $d_2$ are very small,
the deviation is very small, order of magnitude smaller than the ones in Fig. \ref{first}.  
The deviation pattern for sinusoidal square path and sinusoidal cubic path are similar 
to square path. To avoid the overcrowding of figures, we only plotted
the results for square path. 

We finally look at the results for  cubic path, which is the only second-order path among the six. 
From the discussion above, we know that the main deviation of cubic path is of the third order 
because both its $c_2$ and $d_2$  are zero. Therefore, the deviation of cubic path should be
very different from square path. For comparison, the numerical results
for cubic path are plotted in Fig.\ref{cubicp} in a similar fashion as for square path.
There are still oscillations around the second-order analytical result as seen from  
the inset of Fig.\ref{cubicp}(a). However, the oscillations have a very different pattern:
as $c_2$ is zero, the oscillation starts at zero with a much smaller amplitude and frequency. 
In Fig.\ref{cubicp}(b), where the difference is plotted, the oscillations with small oscillations 
are not even visible due to a large kink around $t=500$. This is very different from 
Fig.\ref{second}(b)).  Overall, we see that the deviation of cubic path is much smaller than
square path.

\begin{figure}[htbp]
\centering
\includegraphics[width=1\linewidth]{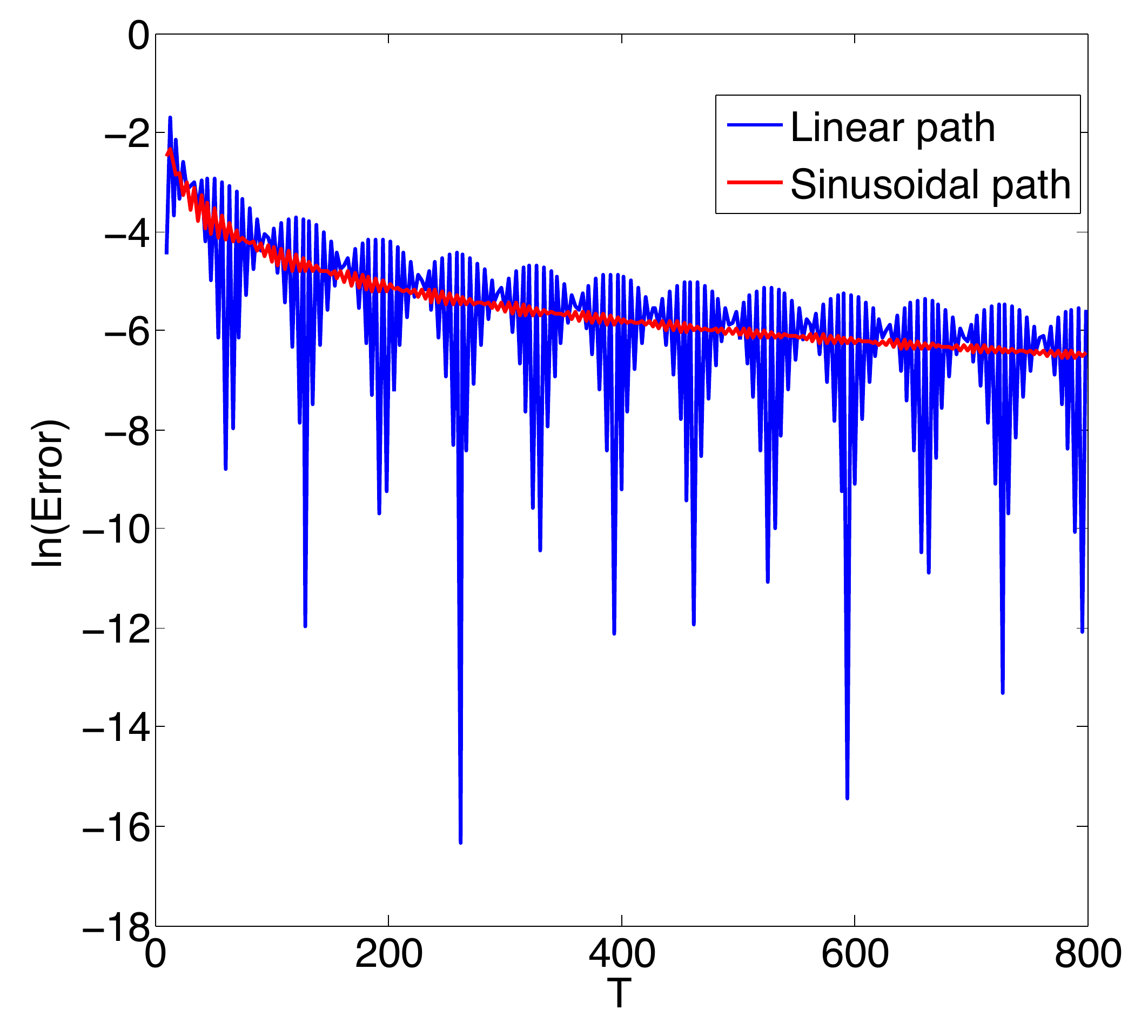}
\caption{(color online) The computation error as a function of computation time $T$ for 
the linear path and the sinusoidal path.  ($N=100$, $M=1$)\label{sectionone}}
\end{figure}
\begin{figure}[htbp]
\centering
\includegraphics[width=1\linewidth]{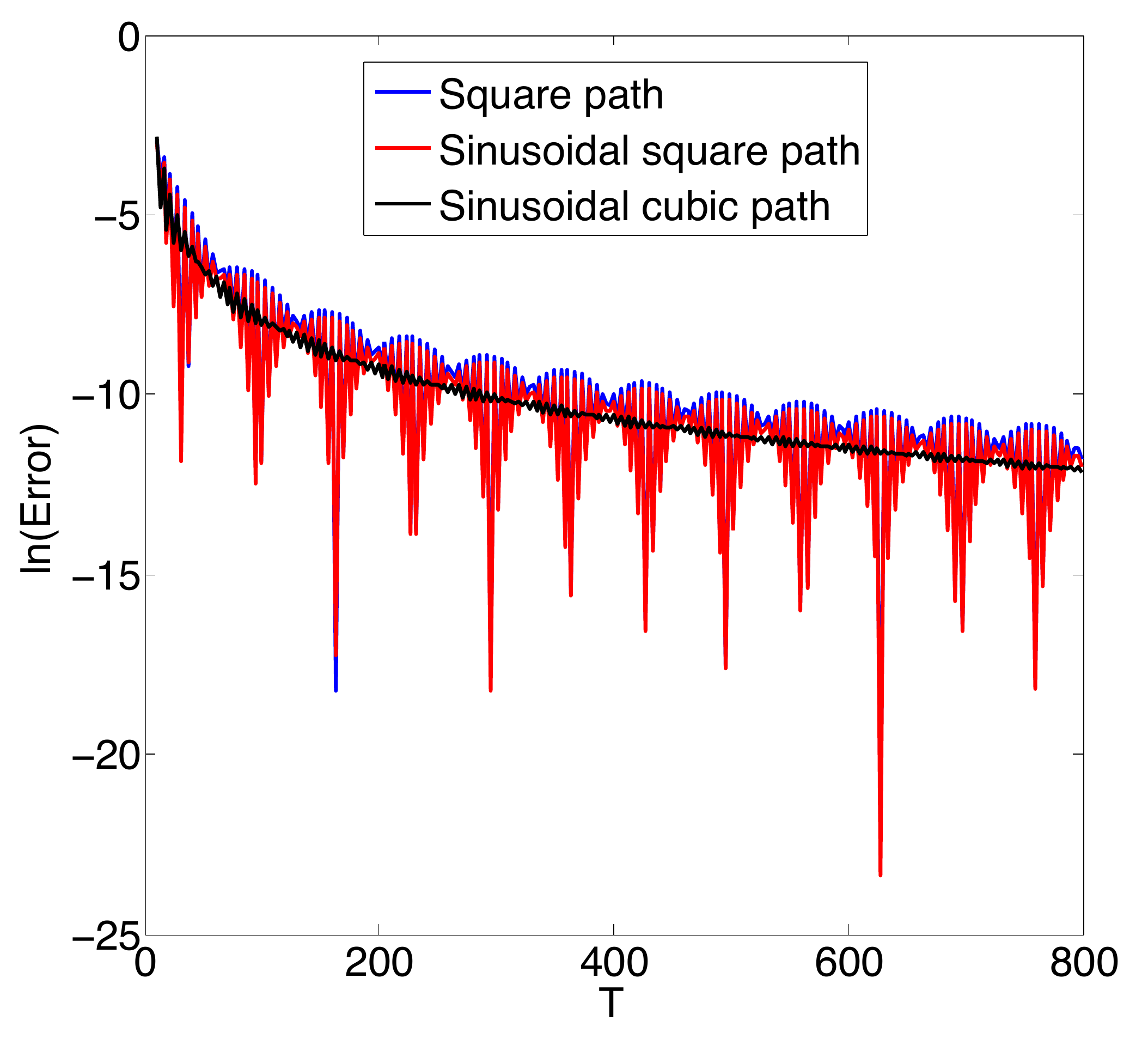}
\caption{(color online) The computation error as a function of computation time $T$ for 
the square path, the sinusoidal square path, and the sinusoidal cubic path.  ($N=100$, $M=1$)\label{sectiontwo}}
\end{figure}
\begin{figure}[htbp]
\centering
\includegraphics[width=1\linewidth]{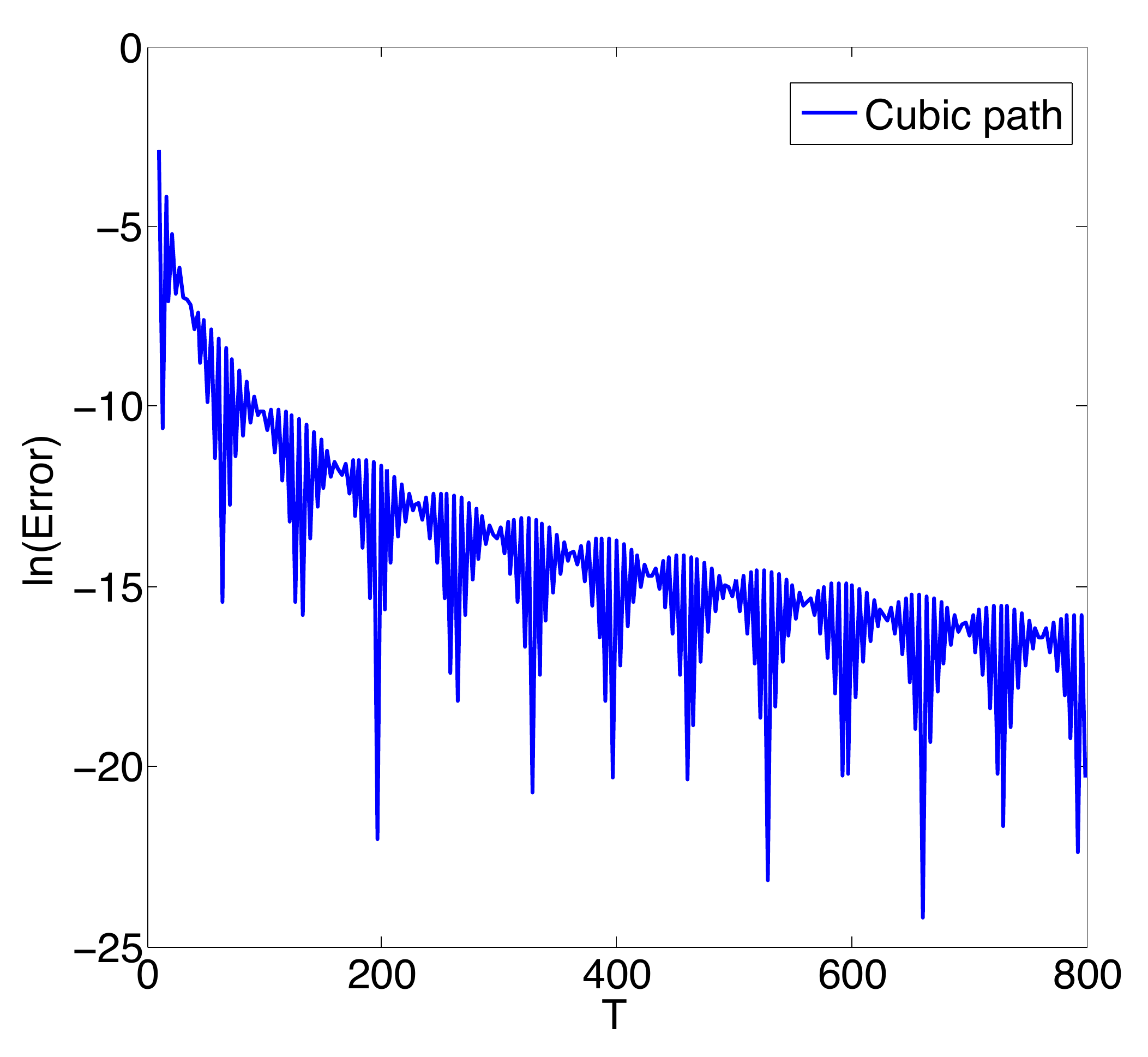}
\caption{(color online) The computation error as a function of computation time $T$ for 
the  cubic path.  ($N=100$, $M=1$)\label{sectionthree}}
\end{figure}

\begin{figure}[htbp]
\centering
\includegraphics[width=0.95\linewidth]{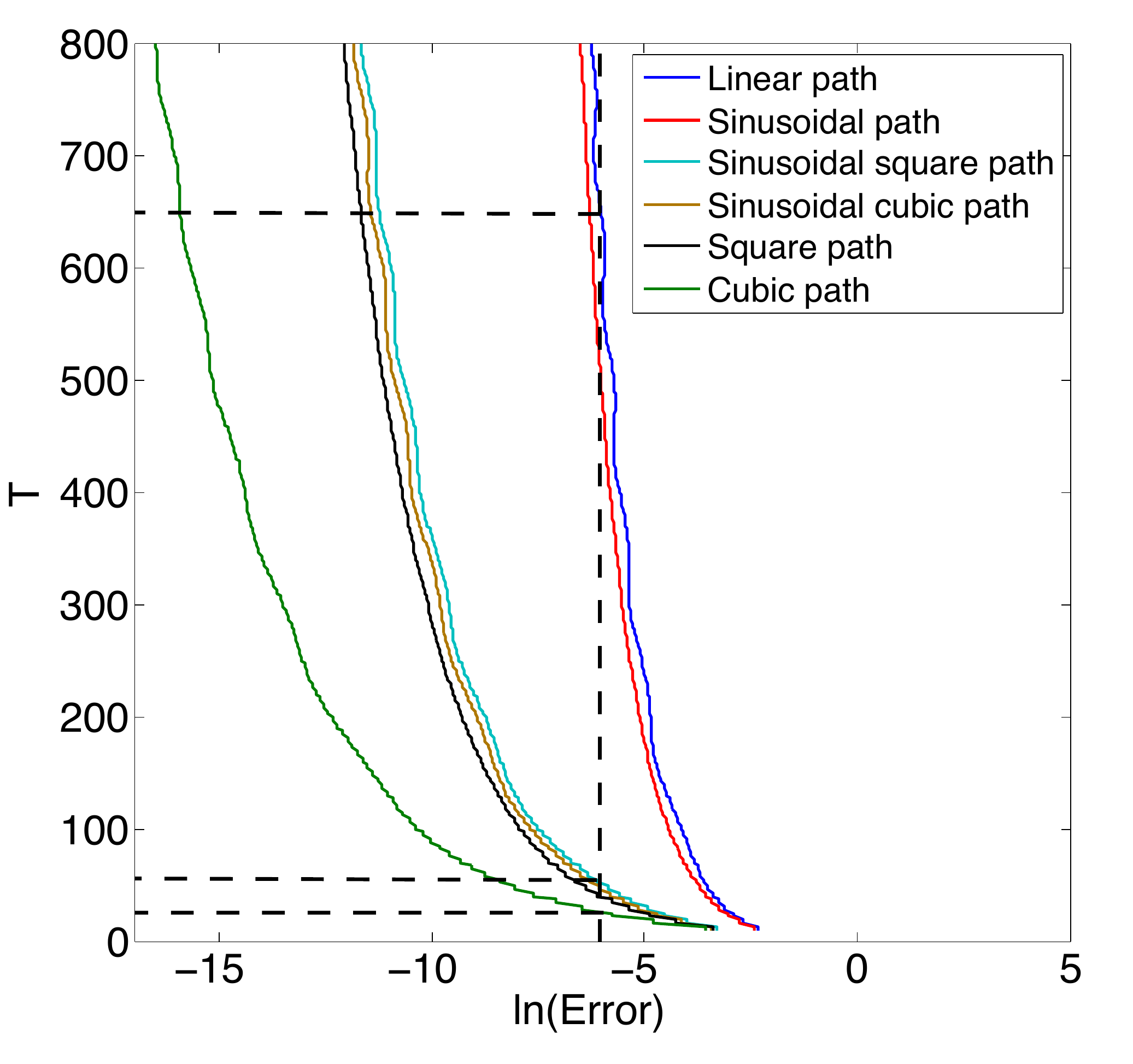}
\caption{(color online)The smoothed computational error with total evolution time of six paths. (N=100, M=1)\label{errorwithtime}}
\end{figure}
For a given total computation time $T$, the success of an algorithm depends 
on the computational error.
The smaller the error the better the algorithm. The computational errors for all the six
proposed paths are computed and plotted against the computation time $T$ in 
Figs. \ref{sectionone},\ref{sectiontwo},\&\ref{sectionthree}. 
Fig. \ref{sectionone} shows the results for the two path of the zeroth order.  The results
for the three first-order paths are shown in Fig.\ref{sectiontwo}.  The cubic path is
of the second order and has the smallest computational error as seen in Fig. \ref{sectionthree}. 
The oscillations in these three figure are originated from the oscillations in 
Figs. \ref{first}\&\ref{second};  they  are not essential. 

We can view the results in Figs. \ref{sectionone},\ref{sectiontwo},\&\ref{sectionthree}
in a different angle: for a given allowed computational error, which path has the 
shortest computation time?  For this angle, we have averaged out
these oscillations and combined the results in 
Figs. \ref{sectionone},\ref{sectiontwo}\&\ref{sectionthree} into Fig. \ref{errorwithtime}.  
In Fig. \ref{errorwithtime}, we see clearly
that the relation between the computation time $T$ and the computational errors 
differs slightly for paths of the same order. However, for
path of different orders, the relation is very different: 
for a certain allowed computational error, the computation time $T$ can differ by 
orders of magnitude; similarly, for a given computation time $T$, 
the error can differ by orders of magnitude. 
For example, at $T=100$, the computational
error for cubic path is almost seven orders of magnitude smaller than the popular linear path. 
If we want to limit the computational error to $10^{-6}$, it only takes $T = 25$ for cubic path; for sinusoidal square path, $T$ is around 46;  for the popular linear path, it would take a much 
longer time $T = 650$. 

Although the oscillations are not essential, they do exhibit interesting patterns. 
The computational errors for four paths, the linear path, the square path, the sinusoidal
square path, and the cubic path, have very similar oscillation patterns while the other two paths, 
the sinusoidal path and the sinusoidal cubic path, share another oscillation pattern. The pattern is largely depend on the relationship between $|c_n|$ and $|d_n|$, the derivatives of $s$ at the beginning and end. Take zeroth-order path as an example. If $|c_1|=|d_1|$,  the actual trajectory can  intersect the adiabatic trajectory for some values of $T$. At those $T$, the computational error is extremely small, as shown in Fig. 1(a). However, if $|c_1|\neq|d_1|$, the actual trajectory will oscillate with time but 
will not intersect the adiabatic trajectory. As a result, the computational error will have some oscillations but never reach zero.

Note that in this section  we have chosen $N = 10, ~M = 1$ 
for Figs. \ref{first},\ref{second}\&\ref{cubicp} just for the clarity of the figures. The essential conclusions drawn from these figures
are the same for larger $N$. \\

\section{Conclusion}
In sum, with the aid of hierarchical theory\cite{Qi}, we have shown that 
it is very effective to shrink the computational error 
by controlling the time derivative of $s(t)$ at the beginning and end of the evolution. 
Our numerical results with six typical adiabatic paths show that a 
path of higher orders (smoother path by intuition) leads to errors of orders  of magnitude smaller. 
Or, for an allowed computational error, the algorithm with a higher-order path
can be order-of-magnitude faster.  The large deviation from the ground state in the middle of the evolution 
is not essential as long as it does not break the 
adiabaticity too much. Although we have focused on quantum search, 
our results and method are general and can be applied to other quantum adiabatic algorithms.

This work is supported by the NBRP of China (2013CB921903,2012CB921300) and
the NSF of China (11274024,11334001,11429402).

\appendix
\section*{Details on first order deviation}
The first order deviation from instantaneous fixed points $(\bar{p}, \bar{q})$ can be written as 
\begin{equation}
	p(t) = \bar{p}[s(t)] + \delta p,~~q(t) = \bar{q}[s(t)]+\delta q.
\end{equation}
  First we consider $s$ is fixed. Using Hamilton equations of motion and Talyor expansion to the first order of $(\delta p, \delta q)$, we have
  \begin{equation}
  	\begin{pmatrix}
  		\dfrac{dp}{dt}\\ \\ \dfrac{dq}{dt}
  	\end{pmatrix}=\Gamma_{0}\begin{pmatrix}
  		\delta p\\ \\ \delta q
  	\end{pmatrix},
  \end{equation}
  where
 \begin{equation}
 	\Gamma_{0} = \begin{pmatrix}
 		-\dfrac{\partial^2 \widetilde{H}_{c}}{\partial q \partial p}
 		& -\dfrac{\partial^2 \widetilde{H}_{c}}{\partial q\partial q}\\
 		\\
 		\dfrac{\partial^2 \widetilde{H}_{c}}{\partial p\partial p}
 		& \dfrac{\partial^2 \widetilde{H}_{c}}{\partial p \partial q}
 	\end{pmatrix}_{p = \bar{p},q = \bar{q}}.
 \end{equation}
 The reason why first-order derivatives of $\widetilde{H}_{c}$ do not appear on the right hand side of Eq.(23) is simply because $(\bar{p},\bar{q})$ is the fixed point.\\
 
 Now we consider $s(t)$ changes slowly with time and examine the dynamics of $(\delta p, \delta q)$. We have
 \begin{equation}
 	\begin{array}{l}
 		\dfrac{dp}{dt} = \dfrac{\partial \bar{p}}{ds}\dot{s}+\dfrac{d\delta p}{dt}\\
 		\\
 		\dfrac{dq}{dt} = \dfrac{\partial \bar{q}}{\partial s}\dot{s}+\dfrac{d\delta q}{dt}.
 	\end{array}
 \end{equation}
  Eq.(23) becomes 
  \begin{equation}
  	\begin{pmatrix}
  		\dfrac{d\delta p}{dt}\\
  		\\
  		\dfrac{d\delta q}{dt}
  	\end{pmatrix}
  	 = \Gamma_{0}(s)\begin{bmatrix}\begin{pmatrix}
  	 	\delta p\\
  	 	\\
  	 	\delta q
  	 \end{pmatrix}-\Gamma_{0}^{-1}(s)\begin{pmatrix}
  	 	\dfrac{\partial \bar{p}}{\partial s}\\
  	 	\\
  	 	\dfrac{\partial \bar{q}}{\partial s}
  	 \end{pmatrix}\dot{s}\end{bmatrix}.
  \end{equation}
  \newline
  
  It can be shown that the determinant $|\Gamma_{0}|$ does not vanish as long as the energy levels of $\widetilde{H}_{s}$ are non degenerate. Previous work\cite{Qi,Fu1,Fu2} shows that $(\delta p, \delta q)$ are a canonical pair and Eq.(26) can be derived from following Hamiltonian,
  \begin{equation}
  \begin{array}{l}
  	H_{1}(s,\dot{s}) = \dfrac{1}{2}(\dfrac{\partial^2 \widetilde{H}_{c}}{\partial q^2})_{\bar{p},\bar{q}}(\delta q - B_{1})^2+\\
  	\\
  	(\dfrac{\partial^2 \widetilde{H}_{c}}{\partial q \partial p})_{\bar{q},\bar{q}}(\delta q -B_{1})(\delta p - A_{1})+\\
  	\\
  	\dfrac{1}{2}(\dfrac{\partial^2 \widetilde{H}_{c}}{\partial p^2})_{\bar{p},\bar{q}}(\delta p - A_{1})^2
  	  \end{array},
  \end{equation}
  where ($A_{1}$,$B_{1}$) is the center of first order deviation $(\delta p, \delta q)$ and defined as 
  \begin{equation}
  	\begin{pmatrix}
  		A_{1}\\
  		\\
  		B_{1}
  	\end{pmatrix}=\Gamma_{0}^{-1}(s)\begin{pmatrix}
  		\dfrac{\partial \bar{p}}{\partial s}\\
  		\\
  		\dfrac{\partial \bar{q}}{\partial s}
  	\end{pmatrix}\dot{s}.
  \end{equation}
  Higher order derivations of ($p,q$) can be derived in similar way.


\begin{thebibliography}{99}
\bibitem{Farhi0}E. Farhi, J. Goldstone, S. Gutmann, and M. Sipser, arXiv:quant-ph/0001106v1 (2000). 
\bibitem{QuantumComputationBook} M. A. Nielesn and I. L. Chuang, \emph{Quantum Computation and Quantum Information} (Cambridge University Press, 2000).
\bibitem{Roland}J. Roland and N. J. Cerf, Phys. Rev. A \textbf{65}, 042308 (2002). 
\bibitem{RezakhaniPRL}A. T. Rezakhani, W.-J. Kuo, A. Hamma, D. A. Lidar, and P. Zanardi, Phys. Rev. Lett. \textbf{103}, 080502 (2009).
\bibitem{Grover}L. K. Grover, Phys. Rev. Lett. \textbf{79}, 325 (1997).
\bibitem{Farhi2}E. Farhi, J. Goldstone, S. Gutmann, J. Lapan, A. Lundgren, and D. Preda, Science \textbf{292}, 472 (2001).
\bibitem{Zhuang}Q. Zhuang, Phys. Rev. A \textbf{90}, 052317 (2014).
\bibitem{Preskill}A. M. Childs, E. Farhi, and J. Preskill, Phys. Rev. A \textbf{65}, 012322 (2001).
\bibitem{Seth}W. M. Kaminsky and S. Lloyd, arXiv:quant-ph/0211152v1 (2002).
\bibitem{DWAVE}http://www.dwavesys.com/\\
sites/default/files/Map\%20Coloring\%20WP2.pdf.


\bibitem{Rezakhani}A. T. Rezakhani, A. K. Pimachev, and D. A. Lidar, Phys. Rev. A \textbf{82}, 052305 (2010).
\bibitem{Qi}Q. Zhang, J. Gong, and B.  Wu, New J. Phys. \textbf{16}, 123024 (2014). 

\bibitem{Lidar}Daniel A. Lidar, Ali T. Rezakhani, and Alioscia Hamma, J. Math. Phys. \textbf{50}, 102106 (2009).
\bibitem{BornFock}M. Born and V. Fock, Z. Phys \textbf{51}, 165 (1928).

\bibitem{Fu1}J. Liu and L. B. Fu, Phys. Rev. A \textbf{81}, 052112 (2010).
\bibitem{Fu2}L. B. Fu and J. Liu, Ann. Phys., N.Y. \textbf{325}, 2425 (2010).
\bibitem{Niu}Jie Liu, Biao Wu, and Qian Niu, Phys. Rev. Lett \textbf{90}, 170404 (2003).

\bibitem{VDam2}A. M. Childs and W. van Dam, Rev. Mod. Phys. \textbf{82}, 1 (2010).
\bibitem{Dirac}P. A. M. Dirac, Proc. R. Soc. \textbf{107}, 725 (1925).



%
\end{thebibliography}
\end{document}